\documentclass[journal]{IEEEtran}

\usepackage{amsmath}
\usepackage{amsfonts}
\usepackage{amssymb}
\usepackage{mathrsfs}
\usepackage{dsfont}
\usepackage{bbm}
\usepackage{bm}
\usepackage{color}
\usepackage{hyperref}
\usepackage{booktabs}
\usepackage{graphicx}
\usepackage{epstopdf}
\usepackage{epsfig}
\usepackage{cite}
\usepackage{enumerate}
\usepackage{amsopn}
\usepackage{ntheorem}
\usepackage{algorithm,algpseudocode}

\newtheorem{theorem}{Theorem}
\newtheorem{remark}{Remark}
\newtheorem{lemma}{Lemma}
\newtheorem{proposition}{Proposition}
\newtheorem{example}{Example}
\newtheorem{corollary}{Corollary}
\newtheorem{definition}{Definition}
\newtheorem{problem}{Problem}
\newtheorem{assumption}{Assumption}

\DeclareMathOperator{\Tr}{Tr}
\DeclareMathOperator*{\argmin}{arg\,min}

\makeatletter

\newcommand{\Rmnum}[1]{\expandafter\@slowromancap\romannumeral #1@}
\makeatother

\title{Max-Min Fair Sensor Scheduling: Game-theoretic Perspective and Algorithmic Solution}

\author{Shuang Wu$^{1}$, Xiaoqiang Ren$^{2}$, Yiguang Hong$^{3}$ and Ling Shi$^{1}$
	\thanks{$^{1}$Shuang Wu and Ling Shi are with Department of Electronic and Computer Engineering, Hong Kong University of Science and Technology, Clear Water Bay, Kowloon, Hong Kong {\tt\small {swuak},{eesling}@ust.hk}}
	\thanks{$^{2}$Xiaoqiang Ren is with ACCESS Linnaeus Center, School of Electrical Engineering, KTH Royal Institute of Technology, 114 28 Stockholm, Sweden {\tt\small {xiaren}@kth.se}}
	\thanks{$^{3}$Yiguang Hong is with Key Laboratory of Systems and Control, Institute of Systems Science, Chinese Academy of Sciences, Beijing 100080, China {\tt\small {yghong}@iss.ac.cn}}
}

\begin{document}

\maketitle

%%%%%%%%%%%%%%%%%%%%%%%%%%%%%%%%%%%%%%%%%%%%%%%%%%%%%%%%%%%%%%%%%%%%%%%%%%%%%%%%%%%
\begin{abstract}
We consider the design of a fair sensor schedule for a number of sensors monitoring different linear time-invariant processes. The largest average remote estimation error among all processes is to be minimized. We first consider a general setup for the max-min fair allocation problem. By reformulating the problem as its equivalent form, we transform the fair resource allocation problem into a zero-sum game between a ``judge" and a resource allocator. We propose an equilibrium seeking procedure and show that there exists a unique Nash equilibrium in pure strategy for this game. We then apply the result to the sensor scheduling problem and show that the max-min fair sensor scheduling policy can be achieved.
\end{abstract}


%%%%%%%%%%%%%%%%%%%%%%%%%%%%%%%%%%%%%%%%%%%%%%%%%%%%%%%%%%%%%%%%%%%%%%%%%%%%%%%%%%
\section{Introduction}
A wireless sensor network consists of a group of sensor nodes deployed in an area. The sensors sense and take measurements of the surrounding environment and transmit the obtained data through a wireless communication channel to a data aggregation center which performs state estimation. Thanks to their large-scale deployment, the wireless sensor networks have gained popularity in a wide range of applications such as environmental monitoring and target tracking~\cite{yick2008wireless}. However, resource constraints, including the amount of energy and the wireless communication channel bandwidth, pose challenges to system design and synthesis.

The sensor scheduling problem has been proposed to utilize the limited resources for enhancing the remote estimation quality. Energy consumption minimization along with sensor lifetime maximization have been studied  in~\cite{bitar2009energy,mo2011sensor,shi2011sensor}. In the mean time, another stream of research has been focusing on balancing the communication burden and a certain cost function related to estimation accuracy. Shi et al.~\cite{shi2011optimal} showed that the optimal schedule for a single sensor scheduling is periodic if packet drops are ignored. Chakravorty and Mahajan~\cite{chakravorty2017fundamental} obtained a general result which stated that the optimal policy is of threshold type. Similar threshold-type structural results are obtained for multiple sensors~\cite{molin2014price,gatsis2015opportunistic,leong2017sensor,ren2018attack}. The cost function in the above mentioned works is the summation of all the estimation errors. In some scenarios, one is not interested in optimizing a collective quantity of each individual, e.g., summation of all the estimation errors, but desires that each individual's performance is above certain threshold. Moreover, each individual has its own interests and may not be willing to reach a collective goal by sacrificing their own benefits. These demands lead to a requirement of fairness in scheduling policies.

In this work, we consider how to fairly allocate the limited communication resources to the sensors. To the best of our knowledge, fairness issue in the sensor scheduling problem has not been addressed. One major challenge is to propose a fairness metric for analysis. The fairness has been studied in operation research and communication community. The works in~\cite{moulin2002fair,correa2007fast,jiang2018towards} studied fairness issue in queuing systems, while the works in~\cite{wang2004distributed,karipidis2008quality,sadeghi2018max} addressed fairness issue for communication protocols. Lan et al.~\cite{lan2010axiomatic} summarized a variety of proposed fairness metrics and developed a set of axioms to identify one metric to be fair. Bertsimas et al.~\cite{bertsimas2011price} pointed out that \emph{max-min} fairness and \emph{proportional} fairness are two axiomatically justified and well accepted notations of fairness. The proportional fairness can be obtained by maximizing the summation of the $\log$ function of the utility of each user. Analysis and computation of solutions to proportional fairness can be done by using the same method as those for multiple sensor scheduling in the literature.

The max-min fairness is taken as the notation of fairness in this work. We consider $n$ sensors measuring $n$ independent linear time-invariant dynamic processes. The sensors transmit their measurements to a remote state estimator through a shared wireless communication channel. Due to the constrained bandwidth, the allowable transmissions are limited. We want the maximal average estimation error covariance among the $n$ sensors to be as small as possible, which constitutes the max-min fairness problem. Interestingly, if the processes are all unstable, the max-min fairness formulation leads to an equal estimation performance (\textbf{Proposition~\ref{proposition: equal performance}}). Although the max-min fairness metric is clearly defined, the corresponding mathematical problem is to be formulated.

There are noticeable amount of works addressing max-min fair scheduling in wireless sensor networks. However, the problems they studied considered are different from ours. The works in~\cite{huang2001max,tassiulas2002maxmin} considered max-min fairness of data flows in the communication links. Their design goal is to maximize the bottleneck of the whole network. The works in~\cite{sridharan2004max,kim2017multi} considered how to schedule transmissions to maximize the lowest throughput over the MAC-layer. In summary, these works considered the fairness solely either from the data link or the routing perspective. None of the existing works addressed the max-min fairness by jointly considering the remote state estimation and bandwidth allocation.

The main results of this paper are summarized as follows.
\begin{enumerate}
	\item We consider the max-min fairness in terms of remote state estimation performance for sensor scheduling problem, which cannot be directly solved using existing numerical schemes. We formulate it as a general max-min fair resource allocation problem. We reformulate it as a two-player zero-sum game, which consists of two players, a ``judge" who adjusts the weights of each sensor and an ``allocator" who determines the resource allocation. We prove that the game possesses pure-strategy Nash Equilibriums (\textbf{Theorem~\ref{theorem: uniqueness of NE}}).
	\item For the general max-min fair allocation problem, we propose an algorithm to seek the equilibrium and show that the algorithm converges to the max-min fair allocation scheme (\textbf{Theorem~\ref{theorem: equilibrium and optimality}} and~\textbf{\ref{theorem: convergence to unique point}}). The max-min fair sensor scheduling problem does not satisfy some of the assumptions made for the general case. We adjust the algorithm as \textbf{Algorithm~\ref{alg:framework}} to seek a max-min fair allocation for sensor scheduling. This adjusted algorithm converges to the max-min allocation with a high accuracy. (\textbf{Theorem~\ref{theorem: fair policy unique and convergent}}).
%	\item We show that how to apply the above methods along with the results from single sensor scheduling to the fair sensor scheduling problem.
%	\item We lastly design an algorithm to recover a strictly feasible scheduling policy from the Nash equilibrium strategy.
\end{enumerate}

The remainder of this paper is organized as follows. In section~\Rmnum{2}, we develop a general framework for the max-min fair resource allocation problem. In section~\Rmnum{3}, we apply the theory developed to solve a fair sensor scheduling problem. The paper is summarized in section~\Rmnum{4}.

\emph{Notations}: The $n$-dimension Euclidean space is denoted by $\mathbb{R}^n$. The bold symbol letter stands for a vector which aggregates all its components, e.g., $\boldsymbol{x}=[x_1,\dots,x_n]^{\top}$. In particular, $\bm{1}=[1,\dots,1]^{\top}$. The inequality between a vector in the Euclidean space means that the inequality holds elementwise. For a matrix $X$, $X^{\top}$ and $\Tr(X)$ stand for the matrix transpose and the trace of the matrix. The operation $P_{\mathcal{X}}(\bm{x})$ denotes the projection of vector $\bm{x}$ into the constrained set $\mathcal{X}$, i.e.,
$P_{\mathcal{X}}(\bm{x}) = \argmin_{\bm{x'}\in\mathcal{X}} \|\bm{x}-\bm{x}'\|_2$,
where $\|\cdot\|_2$ stands for the Euclidean-norm. The probability and condition probability are denoted by $\mathtt{Pr}(\cdot)$ and $\mathtt{Pr}(\cdot | \cdot)$, respectively. The expectation of a random variable is $\mathbb{E}[\cdot]$.

%%%%%%%%%%%%%%%%%%%%%%%%%%%%%%%%%%%%%%%%%%%%%%%%%%%%%%%%%%%%%%%%%%%%%%%%%%%%%%%%%%
\section{The Fair Resource Allocation Problem}

\subsection{Max-Min Fair Resource Allocation}
Consider a group of $n$ agents. Each agent $i=1,\dots,n$ aims to minimize a certain cost $J_i(r_i)$ by consuming certain resource $r_i$. The total resource is limited and characterized by $\sum_{i=1}^n r_i \leq R$. Moreover, the resource utilized by each agent has both an upper bound and a lower bound, i.e., $\underline{r_i} \leq r_i \leq \overline{r_i}$. We aim to find a centralized max-min fair resource utilization scheme $\bm{r}=\begin{bmatrix}r_1,\dots,r_n \end{bmatrix}^\top$ among the $n$ agents as follows.

\begin{problem}\label{problem: resource allocation fairness}
	\begin{align*}
	\min_{\bm{r}} \quad & \max_{i\in\mathcal{I}} J_i(r_i)\\
	\text{s.t.}\quad & \bm{1}^{\top}\bm{r} \leq R,\\
	&\underline{\bm{r}} \leq \bm{r} \leq \overline{\bm{r}},
	\end{align*}
where $\mathcal{I}=\{1,\dots,n\}$.
\end{problem}
This problem aims at optimizing the cost of the ``worst" agent. A central manager solves the above resource allocation problem and each individual uses the allocated resource to optimize its own performance.

\begin{remark}
Problem 1 involves a mixture of numerical (continuous) strategy $\bm{r}$ and categorical (discrete) strategy $i$. When the strategy is either purely continuous or discrete, efficient algorithms can be developed to solve the problem~\cite{nisan2007algorithmic}. If both sets of strategies are discrete, we can reformulate the problem as a linear program  by considering the notion of mixed strategy for discrete strategies. In the linear program setup, the strategy sets are continuous. In a continuous game, we can calculate the gradient of the game value with respect to the strategies, which is useful to seek the Nash equilibrium of the game. However, the mixture of continuous and discrete strategies prohibits direct calculation of the gradients.
\end{remark}

It is expected that more allocated resource leads to a smaller cost and the benefit of using resource should have a decreasing effect. Moreover, $J_i(r_i)$ should be continuous in $r_i$. We therefore made the following assumptions on $J_i(r_i)$.

\begin{assumption}[Continuity]\label{assumption: continuity}
For every $i$, the cost $J_i(r_i)$ is continuous in $r_i$.
\end{assumption}
\begin{assumption}[Strict Monotonicity]\label{assumption: monotonicity between objective and resource}
For every $i$, the cost $J_i(r_i)$ is strictly monotone decreasing in $r_i$, i.e.,
\begin{align*}
J_i(r'_i)>J_i(r_i)
\end{align*}
for any $r'_i<r_i$.
\end{assumption}
\begin{assumption}[Convexity]
For every $i$, the cost $J_i(r_i)$ is convex in $r_i$, i.e.,
\begin{align*}
t J_i(r'_i)+ (1-t)J_i(r_i) \geq J_i(t r'_i + (1-t)r_i)
\end{align*}
for any $r_i,r'_i$ and $t\in[0,1]$.
\end{assumption}
Moreover, we assume the costs are always strictly positive.
\begin{assumption}[Strict Positivity]\label{assumption: strict positivity of costs}
The cost of each agent $J_i(r_i)>0$ for $r_i\in[\underline{r_i},\overline{r_i}]$ and any $i\in\mathcal{I}$.
\end{assumption}

\begin{remark}
The strict monotonicity and the convexity assumption of the cost with respect to allocated resources can be interpreted as follows. Consider the following total cost minimization problem.
\begin{align*}
\min_x \quad &f_i(x)\\
\mathrm{s.t.} \quad &g_i(x)\leq r_i.
\end{align*}
The cost $f_i(x)$ and utilized resource $g_i(x)$ are negatively correlated, i.e., $f_i(x)>f_i(x')$ if and only if $g_i(x)<g_i(x')$ for any $x\neq x'$. Hence, minimization is attained only if $g_i(x)=r_i$. Let $J_i(r_i)=\min_{x:g_i(x)\leq r_i} f_i(x)$. The strict monotonicity of $J_i(r_i)$ then follows. We further assume that strong duality holds, i.e.,
\begin{align*}
\min_x \max_{\lambda\geq0} \; f_i(x)+\lambda (g_i(x)-r_i) = \max_{\lambda\geq0} \min_x \; f_i(x)+\lambda (g_i(x)-r_i).
\end{align*}
Let $x^\star$ and $\lambda^\star$ be the corresponding minimizer and maximizer.
Note that
\begin{align*}
J_i(r_i)=&\min_x f_i(x) + \lambda^\star (g_i(x)-r_i) \\
\leq&f_i(x) + \lambda^\star (g_i(x)-r_i)
\end{align*}
for all $x$. Let $x^{'\star}$ be the solution to the minimization with total cost $r'_i$. We can obtain
\begin{align*}
J_i(r_i) \leq &f_i(x^{'\star}) + \lambda^\star( g_i(x^{'\star}) -r_i )\\
=& J_i(r'_i) + \lambda^\star( r'_i -r_i ).
\end{align*}
In summary, for any $r_i$, there exists a $\lambda^\star$ such that
\begin{align}\label{ieq: convexity and monotonicity of each agent}
J_i(r_i) - \lambda^\star(r'_i-r_i) \leq J_i(r'_i)
\end{align}
for all $r'_i$. From the first order condition of a convex function, the cost $J_i(r_i)$ is convex in $r_i$.
\end{remark}

To make the optimal value of Problem~\ref{problem: resource allocation fairness} bounded, we need further assumptions on the cost objective and the total resource. Define the set of feasible allocation schemes as $$\mathcal{R}\triangleq\{\bm{r}:~\bm{1}^{\top}\bm{r} \leq R,~\underline{\bm{r}} \leq \bm{r} \leq \overline{\bm{r}}\}.$$ We further have the following the assumption of the total resource $R$.
\begin{assumption}\label{assumption: feasiblity}
The lower bounds $\underline{r_i}$ satisfies $\sum_{i\in\mathcal{I}}\underline{r_i} < R$. In addition, there exists a feasible allocation scheme $\bm{r}\in\mathcal{R}$ such that $J_i(r_i)<\infty$ for all $i\in\mathcal{I}$.
\end{assumption}
If the first half of the assumption fails,  the feasible region does not contain an interior point. If the other half fails, the total cost under any allocation is unbounded, which leads to the nonexistence of any max-min fair allocation scheme.

\subsection{Game Reformulation}
The formulation in Problem~\ref{problem: resource allocation fairness} does not reveal any inner structures of the problem and cannot help determine whether there exists a solution to the resource allocation problem. In the following, we transform the original allocation problem as a two-player zero-sum game with continuous actions. With this reformulated problem, we can obtain existence of a solution and some properties of such a solution.

The max-min fair allocation problem can be transformed as a two-player zero-sum game. Let $\mathcal{W}\triangleq\{\bm{w}:\bm{1}^{\top}\bm{w}=1,~\bm{w}\geq\bm{0}\}$. We have two players with strategies $\bm{r}\in\mathcal{R}$ and $\bm{w}\in\mathcal{W}$, respectively. Each of the player is solving a coupled optimization problem
\begin{align}\label{eq: game formulation}
\min_{\bm{r}\in\mathcal{R}} g(\bm{w},\bm{r}) ~\text{and}~\max_{\bm{w}\in\mathcal{W}} g(\bm{w},\bm{r}),
\end{align}
where $g(\bm{w},\bm{r})=\sum_{i=1}^n w_iJ_i(r_i)$. This can be viewed as a zero-sum game between a ``judge" with action $\bm{w}$ and a resource allocator with action $\bm{r}$. The ``judge" wants to maximize it by adjusting $\bm{w}$, while the resource allocator wants to minimize it by allocating $\bm{r}$. If this game possesses at least one pure-strategy Nash equilibrium $(\bm{r}^\star,\bm{w}^\star)$, by the definition of a Nash equilibrium, $\bm{r}^\star$ is a solution of the original problem.

From the results in game theory, an equilibrium in the pure strategy exists for this game.
\begin{theorem}\label{theorem: uniqueness of NE}
The two-player zero-sum game in~\eqref{eq: game formulation} has the following properties:
\begin{enumerate}
\item there exists a pure-strategy Nash equilibrium;
\item the value of the game is unique;
\item there may exist multiple pure-strategy equilibrium pairs, but the weight $w$ at equilibriums is unique.
\end{enumerate}
\end{theorem}
\begin{IEEEproof}
The proof is done in three steps:
\begin{enumerate}
	\item there exists a pure strategy equilibrium;
	\item the value of the game is unique;
	\item $w$ at equilibriums is unique.
\end{enumerate}

We first verify the existence. Note that
\begin{itemize}
	\item both $\mathcal{W}$ and $\mathcal{R}$ is compact and convex;
	\item the utility of $\bm{w}$, $g(\bm{w},\bm{r})$, is continuous in $\bm{r}$, and the utility of $\bm{r}$, $-g(\bm{w},\bm{r})$, is continuous in $\bm{w}$;
	\item $g(\bm{w},\bm{r})$ is continuous and linear (thus also concave) in $\bm{w}$, and $-g(\bm{w},\bm{r})$ is continuous and concave in $\bm{r}$.
\end{itemize}
According to~\cite[Theorem 1.2]{fudenberg1991game}, this continuous game possesses at least one pure-strategy Nash equilibrium.

%The uniqueness of the value of the game is well-known in the literatures, e.g.,
We next prove the uniqueness of the value of the game. Suppose both $(\bm{w}^\star,\bm{r}^\star)$ and $(\bm{w}^{\star\star},\bm{r}^{\star\star})$ are equilibrium strategies.

If $g(\bm{w}^\star,\bm{r}^\star) \leq g(\bm{w}^{\star\star},\bm{r}^{\star\star})$, we have
\begin{align*}
g(\bm{w}^{\star\star},\bm{r}^\star) \leq g(\bm{w}^\star,\bm{r}^\star)
\leq g(\bm{w}^{\star\star},\bm{r}^{\star\star}) \leq g(\bm{w}^{\star\star},\bm{r}^\star),
\end{align*}
which leads to 
\begin{align*}
g(\bm{w}^\star,\bm{r}^\star) = g(\bm{w}^{\star\star},\bm{r}^{\star\star}).
\end{align*}

If $g(\bm{w}^\star,\bm{r}^\star) \geq g(\bm{w}^{\star\star},\bm{r}^{\star\star})$, we have
\begin{align*}
g(\bm{w}^\star,\bm{r}^{\star\star}) \geq g(\bm{w}^\star,\bm{r}^\star)
\geq g(\bm{w}^{\star\star},\bm{r}^{\star\star}) \geq g(\bm{w}^\star,\bm{r}^{\star\star}),
\end{align*}
which also leads to 
\begin{align*}
g(\bm{w}^\star,\bm{r}^\star) = g(\bm{w}^{\star\star},\bm{r}^{\star\star}).
\end{align*}

We lastly prove the uniqueness of $w$ at equilibriums. Again assume both $(\bm{w}^\star,\bm{r}^\star)$ and $(\bm{w}^{\star\star},\bm{r}^{\star\star})$ are equilibrium strategies.
Note that
\begin{align*}
g(\bm{w}^{\star},\bm{r}^{\star}) \leq g(\bm{w}^{\star},\bm{r}^{\star\star}) \leq 
g(\bm{w}^{\star\star},\bm{r}^{\star\star}) \\
g(\bm{w}^{\star\star},\bm{r}^{\star\star}) \leq g(\bm{w}^{\star\star},\bm{r}^{\star}) \leq g(\bm{w}^{\star},\bm{r}^{\star}),
\end{align*}
which leads to
\begin{align*}
g(\bm{w}^{\star},\bm{r}^{\star}) = g(\bm{w}^{\star},\bm{r}^{\star\star}) =  g(\bm{w}^{\star\star},\bm{r}^{\star}) = g(\bm{w}^{\star\star},\bm{r}^{\star\star}).
\end{align*}

Since $J_i(r_i)$ is strictly monotone in $r_i$, we have $J_i(r_i^\star)\neq J_i(r_i^{\star\star})$ for $r_i^{\star}\neq r_i^{\star\star}$. Suppose $\bm{w}^\star \neq \bm{w}^{\star\star}$. As $J_i(r_i)$ are strictly greater than zero, we have $g(\bm{w}^{\star},\bm{r}^{\star}) \neq g(\bm{w}^{\star\star},\bm{r}^{\star})$ and $g(\bm{w}^{\star},\bm{r}^{\star\star}) \neq g(\bm{w}^{\star\star},\bm{r}^{\star\star})$, which violates the uniqueness of the value of the game. Therefore, it holds that $\bm{w}^\star=\bm{w}^{\star\star}$.
\end{IEEEproof}

%If $(\bm{r}^\star,\bm{w}^\star)$ is an equilibrium strategy for the game in~\eqref{eq: game formulation}, then $\bm{r}^\star$ solves Problem~\ref{problem: resource allocation fairness}.
%\begin{IEEEproof}
%We prove by contradiction. Suppose $\bm{r}^\star$ is not a solution of Problem~\ref{problem: resource allocation fairness}. There exists $\tilde{\bm{r}}^\star\in\mathcal{R}$ such that $J_j(r_j^\star) > \max_{i\in\mathcal{I}} J_i(\tilde{r}_i^\star)$ for any $j\in\mathcal{I}$. We can obtain
%\begin{align*}
%\max_{\bm{w}\in\mathcal{W}}\sum_{i\in\mathcal{I}} w_i J_i(r^\star_i)  &=\max_{j\in\mathcal{I}}J_j(r_j^\star)\\
%&> \max_{j\in\mathcal{I}}J_j(\tilde{r}_j^\star) = \max_{\bm{w}\in\mathcal{W}}\sum_{i\in\mathcal{I}} w_i J_i(\tilde{r}^\star_i).
%\end{align*}
%Then there exists $\tilde{\bm{w}}\in\mathcal{W}$ such that $(\tilde{\bm{r}},\tilde{\bm{w}})$ is another equilibrium. The value of the game of this equilibrium strategy is different from that of $(\bm{r}^\star,\bm{w}^\star)$. This contradicts the fact that the value of the game is unique. The proof is thus complete.
%\end{IEEEproof}

As the equilibrium in the pure strategy exists, we are able to obtain a solution of Problem~\ref{problem: resource allocation fairness}. In addition, we can observe some properties of a solution of Problem~\ref{problem: resource allocation fairness}. From the following example, we can see that the optimal value and $w$ at equilibriums of the game are unique, but there are multiple max-min allocation strategies.
\begin{example}
Let $J_1(t)=4-t$, $J_2(t)=J_3(t)=1.5-t$ and $\mathcal{R}=\{\bm{r}:\bm{1}^\top\bm{r}\leq1.5,~\bm{0}\leq\bm{r}\leq1 \}$. It is straightforward that any $\bm{r}$ in $\{\bm{r}: r_1=1, r_2+r_3\leq0.5 \}$ is a max-min fair allocation strategy. However, the corresponding weight $w^\star$ is unique as $\bm{w}^\star = \begin{bmatrix}1&0&0\end{bmatrix}^\top$.
\end{example}

The following corollary connects the max-min fairness with the equal performance. In this case, there exists a unique max-min fair allocation strategy.
\begin{corollary}\label{corollary: max achieve}
If $w^\star_i>0$ for each $i\in\mathcal{I}$, then there exists a unique max-min fair allocation strategy $\bm{r}^\star=[r^\star_1 \; \dots \; r^\star_n]^\top$ such that
\begin{align*}
J_i(r_i^\star) = \max_{j\in\mathcal{I}} J_j(r_j^\star).
\end{align*}
\end{corollary}

\begin{IEEEproof}
Let $w^\star$ be the the one at the equilibrium. As $w^\star$ is unique, the value of the game can be represented by
\begin{align*}
\max_{\bm{w}} \quad& \sum_{i=1}^n w_i J_i(r_i^\star)\\
\text{s.t.} \quad&\bm{1}^{\top}\bm{w}=1,~\bm{w}\geq\bm{0}.
\end{align*}
Noting that the constraints are linear, the constraint qualification conditions hold. The corresponding Karush-Kuhn-Tucker (KKT) conditions are
\begin{align*}
&\bm{J}(\bm{r}^\star)-\lambda^\star\bm{1}+\bm{\mu}^\star=0,&\text{(gradient condtion)}\\
&\bm{\mu}^\star\leq 0, &\text{(dual feasiblity)}\\
&\mu_i^\star w_i^\star=0,~i\in\mathcal{I}&\text{(complementary slackness)}
\end{align*}
where $\lambda^\star$ is the dual variable associated with the constraint $\bm{1}^\top\bm{w}^\star=1$ and $\bm{\mu}^\star$ is the dual variable associated with the constraint $\bm{w}^\star\geq0$.

As $\bm{\mu}^\star \leq 0$, we always have $J_i(r_i^\star)\leq \lambda^\star$. Combining the gradient condition and the dual feasibility, we have $J_i(r_i^\star) = \lambda^\star$, if $\mu^\star_i=0$. From the complementary slackness, we know that
$\mu^\star_i=0$ if $w^\star_i>0$. Therefore if $w^\star_i>0$, we can obtain that $J_i(r^\star_i)=\lambda^\star$, which leads to that $J_i(r^\star_i)=\lambda^\star=\max_i J_i(r^\star_i)$. This also proves that $r^\star$ is unique.
\end{IEEEproof}

As a direct consequence of the corollary, for a general $\bm{w}^\star$, we can obtain $J_i(r^\star_i)=\max_{j\in\mathcal{I}} J_j(r^\star_j)$ if $w^\star_i>0$.

\begin{remark}
Note that $w_i>0$ is only sufficient for the $J_i(r_i^\star)=\max_i J_i(r^\star_i)$. However, this equality may still hold when $w_i=0$. Suppose $w^\star_i=0$ for some $i\in\mathcal{I}$, the consumed resource must be $r^\star_i=\underline{r_i}$ since $J_i(r_i)$ is strictly monotone decreasing. Otherwise, the resource available for other agents is smaller and $g(\bm{w},\bm{r})=\sum_{i:w_i\neq0} w_iJ_i(r_i)$ increases. Therefore, if $w_i^\star=0$, we have
\begin{align*}
\max_{r_i}J_i(r_i)=J_i(\underline{r_i})=J_i(r^\star_i) \leq \lambda^\star = \max_j J_j(r^\star_i),
\end{align*}
which still includes the possibility that $J_i(r^\star_i)= \max_j J_j(r^\star_i)$.
\end{remark}

\subsection{Cost-Based Solution Seeking}
The above theorem only ensures the existence of a pure-strategy equilibrium for the corresponding game, but does not provide a computation procedure to obtain such a equilibrium. One may expect to use a best response algorithm. Specifically, we alternate between solving a linear programming and a convex program. The solution of the linear program is the best response of $\bm{w}$ given $\bm{r}$, while the solution of the convex program is the best response of $\bm{r}$ given $\bm{w}$. A best response algorithm, however, does not converge in general as claimed by~\cite[Sec 1.4.3]{nisan2007algorithmic}. In this work, the best response also fails.

We propose a cost-based algorithm to obtain the max-min fair allocation scheme. Note that for a given allocation $\bm{r}$, if $J_i(r_i)$ is too large, we should increase $r_i$. Conversely, if $J_i(r_i)$ is small, it should donate its resource. Based on this observation, we propose the following algorithm:
\begin{align}\label{eq: cost based algorithm}
\bm{r}(t+1) = T(\bm{r}(t)) \triangleq P_{\mathcal{R}}\Big(\bm{r}_t + \varepsilon \bm{J}(\bm{r}(t))\Big).
\end{align}
The equilibrium of~\eqref{eq: cost based algorithm} is a solution of Problem~\ref{problem: resource allocation fairness}.
\begin{theorem}\label{theorem: equilibrium and optimality}
If $\bm{r}^\star$ is the equilibrium  of the discrete-time projected dynamics~\eqref{eq: cost based algorithm}, i.e., $\bm{r}^\star = T(\bm{r}^\star)$, then for any allocation $\bm{r}$
\begin{align*}
\max_{i\in\mathcal{I}} J_i(r^\star_i) \leq \max_{i\in\mathcal{I}} J_i(r_i).
	\end{align*}
\end{theorem}
The proof is technical and we left it in the appendix in the online version~\cite{wu2019max}.

The update inside the projection only requires local information of the cost function, i.e., the value of the cost $J_i(r_i(t))$ at $r_i(t)$. However, the projection still requires knowledge of every $r_i(t)$. We discuss a distributed extension of this algorithm in Appendix-B in the online version~\cite{wu2019max}. To guarantee the convergence of the algorithm in~\eqref{eq: cost based algorithm}, we impose the following assumption on $J_i(r_i)$.
\begin{assumption}\label{assumption: first order lipchitz condition}
The optimal value $J_i(r_i)$ satisfies, for any $r_i$ and $r_i'$ in $[\underline{r_i},\overline{r_i}]$, 
\begin{enumerate}[(1)]
\item(strong monotonicity) there exists a positive real number  $\alpha_i$ such that
	\begin{align*}
	\alpha_i | r_i - r_i' | \leq | J_i(r_i) - J_i(r_i') |;
	\end{align*}
\item($\beta_i$-Lipschitz) there exists a positive real number $\beta_i$ such that
	\begin{align*}
	| J_i(r_i) - J_i(r_i') |  \leq  \beta_i | r_i - r_i' |.
	\end{align*}
\end{enumerate}
\end{assumption}

This assumption means that the variations of the costs with respect to allocated resources are mild.

\begin{theorem}\label{theorem: convergence to unique point}
There exists $\varepsilon \in (0,\frac{2\alpha}{\beta^2})$, where $\alpha=\min_i \alpha_i$ and $\beta = \max_i \beta_i$, such that the algorithm~\eqref{eq: cost based algorithm} converges with a linear convergence rate. 
\end{theorem}
\begin{IEEEproof}
We prove the convergence to a unique point by showing that the update is contractive.
Because of the nonexpansive property of a projection to a convex set, we have
\begin{align*}
\| T(\bm{r}) - T(\bm{r}') \|^2
\leq \| \bm{r} + \varepsilon \bm{J}(\bm{r}) - \bm{r}' - \varepsilon \bm{J}(\bm{r}') \|^2
\end{align*}
Expanding the right hand side, we obtain
\begin{align*}
&\| T(\bm{r}) - T(\bm{r}') \|^2 \\
\leq & \| \bm{r} - \bm{r}' \|^2 + 2\varepsilon (\bm{r} - \bm{r}')^\top (\bm{J}(\bm{r}) - \bm{J}(\bm{r}')) \\
&+ \varepsilon^2\|\bm{J}(\bm{r}) - \bm{J}(\bm{r}')\|^2.
\end{align*}

Note that for each $i$, $(r_i-r'_i)(J_i(r_i)-J_i(r'_i))<0$ due to the monotonicity of $J_i(r_i)$. Therefore, we have
\begin{align*}
(\bm{r} - \bm{r}')^\top (\bm{J}(\bm{r}) - \bm{J}(\bm{r}')) 
= - |\bm{r} - \bm{r}'|^\top|\bm{J}(\bm{r}) - \bm{J}(\bm{r}')|.
\end{align*}

From Assumption~\ref{assumption: first order lipchitz condition}, there exists a $\alpha_i>0$ such that
\begin{align*}
|J_i(r_i) - J_i(r'_i) | \geq \alpha_i |r_i - r'_i|.
\end{align*}
Let $\alpha=\min_i \alpha_i$, we have
\begin{align*}
|\bm{r} - \bm{r}'|^\top|\bm{J}(\bm{r}) - \bm{J}(\bm{r}')| 
\geq  {\alpha}\| \bm{r} - \bm{r}' \|^2,
\end{align*}
which leads to
\begin{align*}
(\bm{r} - \bm{r}')^\top (\bm{J}(\bm{r}) - \bm{J}(\bm{r}')) \leq -{\alpha}\| \bm{r} - \bm{r}' \|^2.
\end{align*}
Consequently,
\begin{align*}
&\| T(\bm{r}) - T(\bm{r}') \|^2 \\
\leq & \| \bm{r} - \bm{r}' \|^2 - 2\varepsilon{\alpha} \|\bm{r} - \bm{r}'\|^2+ \varepsilon^2\|\bm{J}(\bm{r}) - \bm{J}(\bm{r}')\|^2.
\end{align*}

Meanwhile, from Assumption~\ref{assumption: first order lipchitz condition}, we also know that
\begin{align*}
\|\bm{J}(\bm{r}) - \bm{J}(\bm{r}')\|^2 \leq \beta\|\bm{r} - \bm{r}' \|^2,
\end{align*}
where $\beta= \max_i \beta_i$. We can obtain
\begin{align*}
\| T(\bm{r}) - T(\bm{r}') \|^2 \leq  \Big[1+(\varepsilon^2\beta^2-2\varepsilon\alpha)\Big]\| \bm{r} - \bm{r}' \|^2.
\end{align*}
Therefore, as long as $0<\varepsilon<\frac{2\alpha}{\beta^2}$, there exists a real number $0<c<1$ and a point $\bm{r}^\star=T(\bm{r}^\star)$ such that, for any $\bm{r}\in\mathcal{R}$,
\begin{align*}
\|T(\bm{r})-\bm{r}^\star\|^2 = \|T(\bm{r})-T(\bm{r}^\star)\|^2 \leq c \|\bm{r}-\bm{r}^\star\|^2.
\end{align*}
Denoting the error of $\bm{r}(k)$ as $e(k)=\|\bm{r}(k)-\bm{r}^\star\|^2$, we can obtain $e(k+1) \leq c e(k)$, which proves the linear convergence rate.
\end{IEEEproof}

%\subsection{A Perspective From Generalized Nash Equilibrium}
%The max-min fair allocation scheme can also be formulated as generalized Nash Equilibrium problem (GNEP), in which each agent solves the following problem
%\begin{align*}
%\min_{r_i} \quad &J_i(r_i)\\
%\text{s.t.} \quad &r_i + \sum_{j:j\neq i} r_j \leq R\\
%& \underline{r_i} \leq r_i \leq \overline{r_i}.
%\end{align*}
%%%%%%%%%%%%%%%%%%%%%%%%%%%%%%%%%%%%%%%%%%%%%%%%%%%%%%%%%%%%%%%%%%%%%%%%%%%%%%%%%%
\section{Application in Fair Sensor Scheduling}

\subsection{Problem Setup}
Consider $n$ LTI processes, each measured by one sensor. The dynamics are as follows:
\begin{align}
x^{(i)}(k+1) = A_ix^{(i)}(k) + w^{(i)}(k),\label{eq: process dynamics}\\
y^{(i)}(k) = H_i{x}^{(i)}(k) + {v}^{(i)}(k),\label{eq: measurement}
\end{align}
where $i \in N \triangleq \{1,\ldots,n\}$, ${x}^{(i)}(k)\in\mathbb{R}^{n_i}$ is the state of the $i$-th system at time $k$ and ${y}^{(i)}(k)\in\mathbb{R}^{m_i}$ is the noisy measurement taken by the sensors. For all processes and $k\geq0$, the state disturbance noise ${w}^{(i)}(k)$'s, the measurement noise ${v}^{(i)}(k)$'s and the initial state ${x}^{(i)}(0)$'s are mutually independent Gaussian random variables, which follow Gaussian distributions as ${w}^{(i)}(k)\sim\mathcal{N}({0},{Q}_i)$, ${v}^{(i)}(k)\sim\mathcal{N}({0},{R}_i)$ and ${x}^{(i)}(0)\sim\mathcal{N}({0}, {\Pi}_i)$. The covariance matrices ${Q}_i$ and ${\Pi}_i$ are positive semidefinite, and ${R}_i$ is positive definite. We assume that, for every $i \in N$, the pair $({A}_i,{H}_i)$ is observable and the pair $({A}_i,\sqrt{{Q}}_i)$ is controllable.

The sensors measure the process states and compute local state estimates using a Kalman filter. After that, the sensors transmit the estimate data through a communication network under resource constraints to a remote state estimator. The remote state estimator will either synchronize the remote state estimates with the local state estimate if the updated data is received, or use process dynamics to predict the state estimates if no data is received. For sensor $i$ at time $k$, we use $a_i(k)=1$ to denote transmission and $a_i(k)=0$ otherwise. The estimation error covariance of the remote estimator for each process $P^{(i)}$ at time $k+1$ can be computed as follows:
\begin{align*}
P^{(i)}(k+1)=
\begin{cases}
\overline{P}_i, &\text{if~}a_i(k)=1,\\
A_iP^{(i)}(k)A_i^{\top}+Q_i, &\text{if}~a_i(k)=0,
\end{cases}
\end{align*}
where $\overline{P}_i$ is the steady state of the state estimation error covariance of the local Kalman filter.

In this work, we assume that the information available to transmission decision is
\begin{align*}
\tau_i(k) = \min\{t\geq0:a_i(k-t)=1\},
\end{align*}
which is the time elapsed since the last successful transmission. Transmission decisions of the senors are defined by
\begin{align*}
f_i(\tau_i(k)) := \mathtt{Pr}(a_i(k)|\tau_i(k)),
\end{align*}
where $a_i(k)=1$ denotes transmission and $a_i(k)=0$ denotes no transmission. We are interested in allocating the average communication rates of the senors to minimize the largest average remote state estimation error. This can be formulated as
\begin{align}\label{problem: sensor fair problem}
\min_{\bm{r}\in\mathcal{R}} \max_{i\in\mathcal{I}} J_i(r_i),
\end{align}
where the constraint set is
\begin{align*}
\mathcal{R} := \{\bm{r}: \bm{1}^\top\bm{r}\leq R,~\bm{0}\leq\bm{r}\leq 1 \},
\end{align*}
and the performance metric is
\begin{align*}
J_i(r_i):=\limsup_{T\to\infty}\frac{1}{T}\sum_{k=0}^{T-1} \Tr(P^{(i)}(k))).
\end{align*}
To transform the problem as the standard setting in the last section, we need to study properties of $J_i(r_i)$. This requires analysis of optimal scheduling polices for a single sensor.

\begin{remark}
	Some papers studied the state estimation problem from a zero-sum game perspective~\cite{banavar1992game,shen1997game}. Their problems are different from ours. They studied the worst-case estimation error estimator in face of model uncertainty. The related game is between the state estimator and the nature who exerts uncertain disturbance to the system. The estimator wants to minimize the estimation error while the nature wants to maximize the estimation error.
\end{remark}

\subsection{Single Sensor Scheduling}
\begin{figure}[t]
	\centering
	\includegraphics[width=0.3\textwidth]{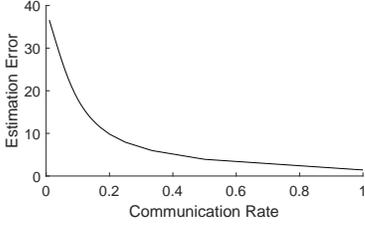}
	\caption{The optimal average estimation error of the single sensor scheduling under constrained communication is piecewise linear, convex and nonincreasing w.r.t. communication rate.}
	\label{fig:single_constrained}
\end{figure}
Given an average communication rate $r_i$ for sensor $i$, the following single sensor scheduling problem is well solved
\begin{align*}
\min_{f_i} \quad& J_i(f_i)\\
\text{s.t.} \quad& C_i(f_i) \leq r_i,
\end{align*}
where $f_i$ is the scheduling policy for sensor $i$ and $C_i(f_i)=\lim_{T\to\infty} \frac{1}{T+1}\mathbb{E}[\sum_{k=0}^{T} a_i(k)]$. The settings in~\cite{chakravorty2017fundamental} are similar to this problem. By using a similar approach, we can obtain the same results for the single sensor scheduling problem as those in~\cite{chakravorty2017fundamental}. We recap essential results of this single sensor scheduling problem.

Recall that $\tau_i(k)$ stands for the time elapsed since the last successful transmission. A threshold policy with threshold $\xi_i$ is defined as follows. Transmit if $\tau_i(k)\geq\xi_i$ and do not transmit otherwise. An optimal policy of the constrained single sensor scheduling is a randomized threshold strategy, which randomizes between two thresholds $\xi_i$ and $\xi_i+1$, i.e.,
\begin{align*}
a_i(k) =
\begin{cases}
0,&\text{ if }\tau_i(k)<\xi_i,\\
0,&with~probability~1-{b_i}, \text{ if }\tau_i(k)=\xi_i,\\
1,&with~probability~{b_i}, \text{ if }\tau_i(k)=\xi_i,\\
1,&\text{if }\tau_i(k)>\xi_i.
\end{cases}
\end{align*}

The communicate rate of the randomized threshold optimal policy is exactly $r_i$. Therefore, given the allocated communication rate $r_i$, we can calculate the corresponding threshold $\xi_i$ and randomization parameter $b_i$ as
\begin{align*}
\xi_i = \lfloor \frac{1}{r_i} - 1 \rfloor,~
b_i =\xi_i+1+\frac{r_i-1}{r_i},
\end{align*}
where $\lfloor\cdot\rfloor$ denotes round downwards.

Given the parameters of the optimal randomized policy, we can compute the corresponding optimal estimation error $J_i(r_i)$. For the optimal single sensor scheduling under constrained communication, the relation of the communication rate and the average estimation error is shown in Fig. \ref{fig:single_constrained}. It can be proven that the optimal estimation error $J_i(r_i)$ is piecewise linear, convex, continuous and strictly decreasing with respect to $r_i$~\cite{chakravorty2017fundamental}.

\subsection{Max-Min Fair Allocation Approach}
\begin{algorithm}
	\caption{Solving the average constraint problem}
	\label{alg:framework}
	\begin{algorithmic}[1]
		\State Initialize $\bm{r}(1)$, $0<\eta<1$, $\varepsilon_r>0$, $\varepsilon>0$, $t=1$ 
		\State Set
		\begin{align*}
		\underline{r_i} \leftarrow 
		\begin{cases}
		0,&\text{if process~}i\text{~is stable},\\
		\eta, &\text{if process~}i\text{~is unstable}.
		\end{cases}
		\end{align*}
		\State \textbf{repeat}
		\State \quad  $\mathcal{R} \leftarrow  \{\bm{r}:~\bm{1}^{\top}\bm{r} \leq R,~\underline{\bm{r}} \leq \bm{r} \leq \bm{1}\}$
		\State \quad \textbf{repeat}
		\State	\qquad	\begin{align*}
		&r_i(t+1) \leftarrow r_i(t) + \varepsilon J_i(r_i(t))\\
		&\bm{r}(t+1) \leftarrow \argmin_{\bm{r}\in\mathcal{R}} \|\bm{r}(t+1) - \bm{r}\|_2^2 \\
		&t \leftarrow t+1\\
		& {\varepsilon} \leftarrow  \frac{1}{\frac{1}{\varepsilon}+1}
		\end{align*}
		\State \quad \textbf{until} $\|\bm{r}(t) - \bm{r}(t-1) \| \leq\varepsilon_r$
		\State \quad $\tilde{\bm{r}}\leftarrow \bm{r}(t)$
		\State \quad \textbf{if} for all unstable processes
		\State	\qquad $\tilde{r}_i>\underline{r_i}$
		\State \qquad \textbf{break}
		\State \quad \textbf{else} for all unstable processes
		\State \qquad  $\underline{r_i} \leftarrow  \eta \cdot \underline{r_i}$ 
		\State \quad \textbf{end if}
		\State \textbf{until} $\tilde{r}_i>\underline{r_i}$ for all unstable processes
	\end{algorithmic}
\end{algorithm}

An optimal allocation for~\eqref{problem: sensor fair problem} can be found by solving the following zero-sum game
$
\max_{\bm{w}\in\mathcal{W}} \min_{\bm{r}\in\mathcal{R}} \sum_{i=1}^n w_i J_i(r_i).
$
Note that Assumption~\ref{assumption: monotonicity between objective and resource}-\ref{assumption: strict positivity of costs} are satisfied from the results in the single sensor scheduling problem. Moreover, the total bandwidth $R$ should be strictly greater than $0$. Then an allocation scheme $r_i=\frac{R}{n}$ suffices to verify Assumption~\ref{assumption: feasiblity}. By applying the existence, we see that there exists at least one fair allocation policy $\bm{r}^\star$.

In addition, this formulation reveals the following result which bridges the max-min fairness and the identical estimation error, which is an extension of Corollary~\ref{corollary: max achieve}.
\begin{proposition}\label{proposition: equal performance}
	The remote estimation errors of the sensors are equal for unstable processes.
\end{proposition}
\begin{IEEEproof}
Note that we must have $w^\star_i>0$ for unstable processes. Otherwise, if $w^\star_i=0$, we have $J_i(\underline{r_i})$ being unbounded, which certainly does not meet the max-min fair requirement. According to Corollary~\ref{corollary: max achieve}, the result follows.
%\textcolor{blue}{Since $w^\star_i>0$ for all unstable processes, from the complementary slackness, we know that $\mu^\star_i=0$. From the gradient condition, we know that $\bm{J}^\star(\bm{r}^\star)=\lambda \bm{1}$, which means that theses processes should have the same value $\max_i J_i(r^\star_i)$.}
\end{IEEEproof}

\begin{remark}
	According to the above proposition, the min-max criterion leads to fairness in the sense of equal remote estimation performance if the processes are unstable. Nevertheless, if there exist stable processes, it could be infeasible to find a scheduling policy such that the remote estimation performances of each process are equal. This occurs if the estimation error of the state prediction for the $i$-th processes is less than the steady state estimation error of the Kalman filter for the $j$-th process, i.e.,
$\Tr(P_i) < \Tr(P_j)$,
	where $P_i=f_p^{(i)}(P_i)$ and $P_j=f_c^{(j)}(f_p^{(j)}(P_j))$ with $f_p^{j}(X)=A_jXA_j^\top+Q_j$ and $f_c^{j}(X)=X-XC_j^\top(C_jXC_j^\top+R_j)^{-1}C_jX$. In this case, we never have $J_i(r_i) = J_j(r_j)$ for any allocation scheme because $\Tr(P_i)$ is the largest estimation error of sensor $i$ while $\Tr(P_j)$ is the lowest estimation error of sensor $j$ .
\end{remark}

To compute the max-min fair allocation scheme using the algorithm in~\eqref{eq: cost based algorithm}, the validity of Assumption~\ref{assumption: first order lipchitz condition} should also be verified. Note that the optimal average estimation performance can be unbounded on $r_i\in(0,1]$, which makes the Lipschitz condition in Assumption~\ref{assumption: first order lipchitz condition} invalid. We overcome this issue by imposing a very small lower bound of the minimum communication rate $r_i\in[\underline{r_i},1]$ for those unstable processes. On the compact interval $[\underline{r_i},1]$ with $\underline{r_i}>0$, the cost objective $J_i(r_i)$ satisfies Assumption~\ref{assumption: first order lipchitz condition}. For any lower bound $\underline{r_i}$, if $r^\star_i=\underline{r_i}$, we set $\underline{r_i}$ to be a smaller value, e.g., $\eta\underline{r_i}$ with $0<\eta<1$, and then run the algorithm~\eqref{eq: cost based algorithm} again. This iterative scheme stops when $r_i^\star>\underline{r_i}$. In practice, however, if the initial $\underline{\bm{r}}$ is small enough, only one iteration for $\underline{\bm{r}}$ is needed. This revised fair allocation scheme seeking algorithm is summarized in Algorithm~\ref{alg:framework}. Essentially, the inner loop corresponds to the original algorithm in~\eqref{eq: cost based algorithm}, and the outer loop seeks the appropriate $\underline{r_i}$.

Note that we use diminishing step sizes in the algorithm. This is proposed to ensure that the step size meets the convergence condition that $\varepsilon\in(0,\frac{2\alpha}{\beta^2})$ in finite steps as mentioned in Theorem~\ref{theorem: convergence to unique point}. This condition requires knowledge of $\alpha$ and $\beta$ before implementing the algorithm. By using the diminishing step sizes, such a requirement is removed. Moreover, the diminishing step sizes are always greater than zero, which means that they also meet the positivity condition of the step sizes in Theorem~\ref{theorem: convergence to unique point}.
The convergence of the revised algorithm is guaranteed in the following theorem.
\begin{theorem}\label{theorem: fair policy unique and convergent}
Algorithm~\ref{alg:framework} converges to $\tilde{\bm{r}}^\star$, which lies in an $\frac{c}{1-c}\varepsilon_r$-neighborhood of a max-min fair allocation policy of~\eqref{problem: sensor fair problem}, i.e., $\|\bm{r}^\star-\tilde{\bm{r}}^\star\|<\frac{c}{1-c}\varepsilon_r$ with $\bm{r}^\star$ being a solution of the max-min fair allocation for~\eqref{problem: sensor fair problem} and $c$ being the contraction constant for the mapping $T(\cdot)$. Moreover, $c=1+(\varepsilon^2\beta^2-2\varepsilon\alpha)$, where $\varepsilon$ is the stepsize when the algorithm terminates, and $\alpha$ and $\beta$ are the same as those in Theorem~\ref{theorem: convergence to unique point}. In addition, for each $\underline{\bm{r}}$, the convergence rate is linear in time.
\end{theorem}
\begin{IEEEproof}
For each $\mathcal{R}$, the iteration~\eqref{eq: cost based algorithm} satisfies the Assumptions required for convergence. The positive decreasing step sizes $\varepsilon$'s satisfy the requirement for the iteration to be a contraction mapping in a finite time. By using the contraction property developed in Theorem~\ref{theorem: convergence to unique point}, the iteration~\eqref{eq: cost based algorithm} converges for each $\mathcal{R}$ and the convergence rate is linear. If $\tilde{r}_i=\underline{r_i}\neq 0$ for any unstable process, there exists a smaller $\underline{r_i}$, which decreases $\tilde{r}_i$ and $\max_j J_j(\tilde{r}_i)$ simultaneously. If $\tilde{r}_i>\underline{r_i}$ for all unstable processes, then $\max_j J_j(\tilde{r}_i)$ cannot decrease by choosing smaller $\underline{r_i}$'s. As $\underline{r}_i$ is lower bounded by zero, the outer loop terminates in a finite time. Finally, let $\bm{r}$ be such that $T(\bm{r})=\tilde{\bm{r}}^\star$. We can observe that
$
\|\bm{r} - \bm{r}^\star\| = \|\bm{r} - T(\bm{r}) + T(\bm{r})- \bm{r}^\star\|
\leq  \varepsilon_r + c \|\bm{r} - \bm{r}^\star \|,
$
which is equivalent to
$
\|\bm{r} - \bm{r}^\star\| \leq \frac{\varepsilon_r}{1-c}.
$
As $\|\tilde{\bm{r}}^\star- \bm{r}^\star\| \leq c\|\bm{r} - \bm{r}^\star \|$, we can obtain $\|\tilde{\bm{r}}^\star- \bm{r}^\star\| \leq \frac{c}{1-c}\varepsilon_r$. This completes the proof.
\end{IEEEproof}

%The computation can be done efficiently. We need to compute $\frac{\partial}{\partial r_i} J_i(r_i)$. As the thresholds are countable, we can compute the value of $J_i(r_i)$ for each threshold policy beforehand and store the values of the slopes in a table. By using the piecewise linear property of the estimation error v.s. communication rate curve (Fig.~\ref{fig:single_constrained}), we can see that $w_i(t)\frac{\partial}{\partial r_i}J_i(r_i(t))$ is the slope multiplied by $w_i(t)$. The two projections are projected onto a linear space, which can also be done efficiently.

%%%%%%%%%%%%%%%%%%%%%%%%%%%%%%%%%%%%%%%%%%%%%%%%%%%%%%%%%%%%%%%%%%%%%%%%%%%%%%%%%%
\section{Numerical Examples}
In this section, we consider five LTI processes. The process parameters in~\eqref{eq: process dynamics} and~\eqref{eq: measurement} are as follows:
\begin{align*}
&{A}_1=\begin{bmatrix}1.2 &0 \\ 0 &0\end{bmatrix}, &{Q}_1&=\begin{bmatrix}4 &0 \\ 0 &1\end{bmatrix};\\
&{A}_2=\begin{bmatrix}1.1 &1 \\ 0 &1\end{bmatrix},  &{Q}_2&=\begin{bmatrix}1 &0 \\ 0 &4\end{bmatrix};\\
&{A}_3=\begin{bmatrix}1.2 &1 \\ 0 &0.8\end{bmatrix}, &{Q}_3&=\begin{bmatrix}1 &0 \\ 0 &4\end{bmatrix}\\
&{A}_4=\begin{bmatrix}0.8 &0.6 \\ 0 &0.9\end{bmatrix}, &{Q}_4&=\begin{bmatrix}16 &0 \\ 0 &1\end{bmatrix};\\
&{A}_5=\begin{bmatrix}0.3 &1 \\ 0 &0.1\end{bmatrix}, &{Q}_5&=\begin{bmatrix}0.3 &0 \\ 0 &1.2\end{bmatrix},
\end{align*}
and $C_i$ and $R_i$ are two-dimensional identity matrix for all $i=1,2,3,4,5$.

\begin{figure}[t]
	\centering
	\includegraphics[width=0.35\textwidth]{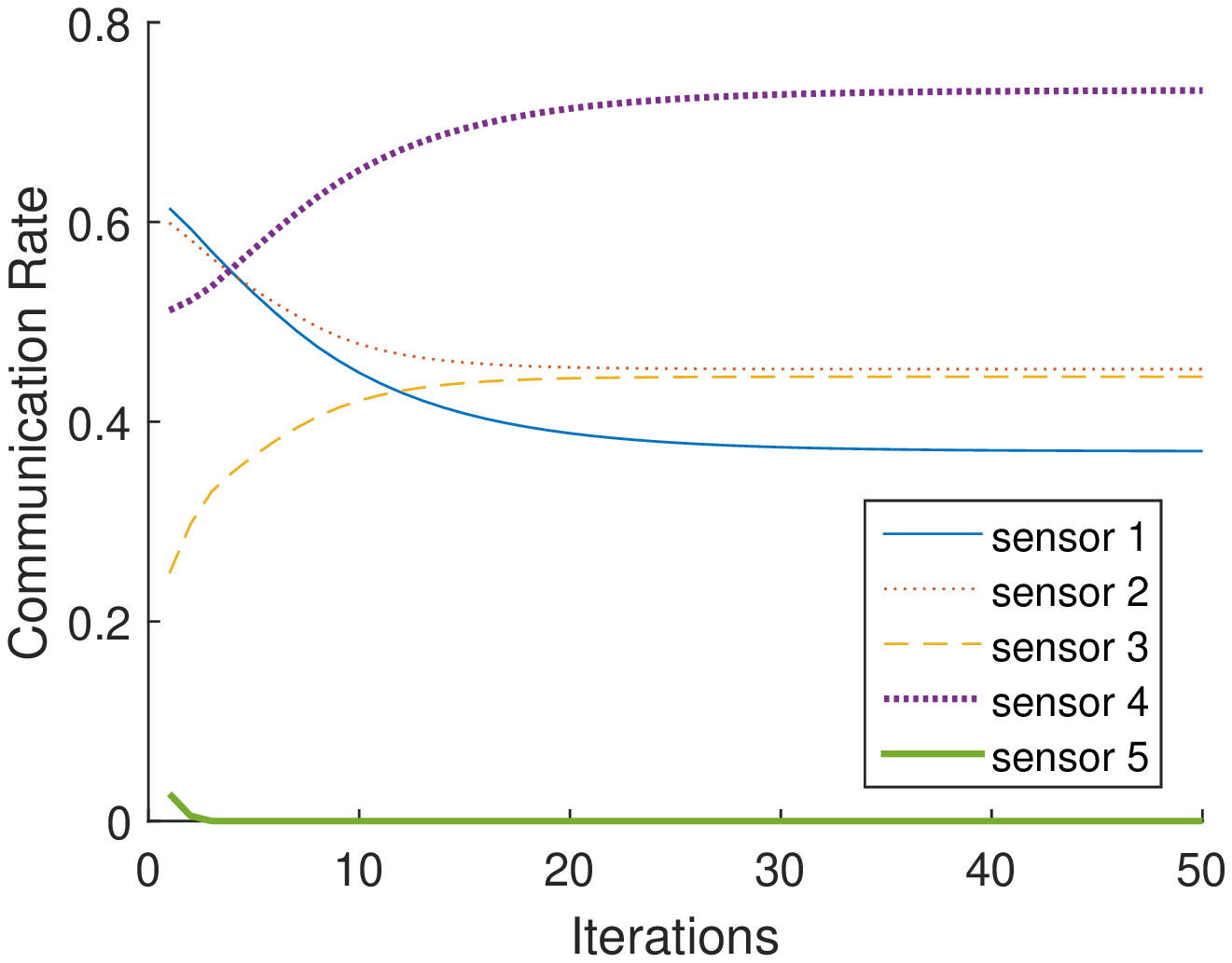}
	\caption{The bandwidth allocation of each processes.}
	\label{fig:2}
	\includegraphics[width=0.35\textwidth]{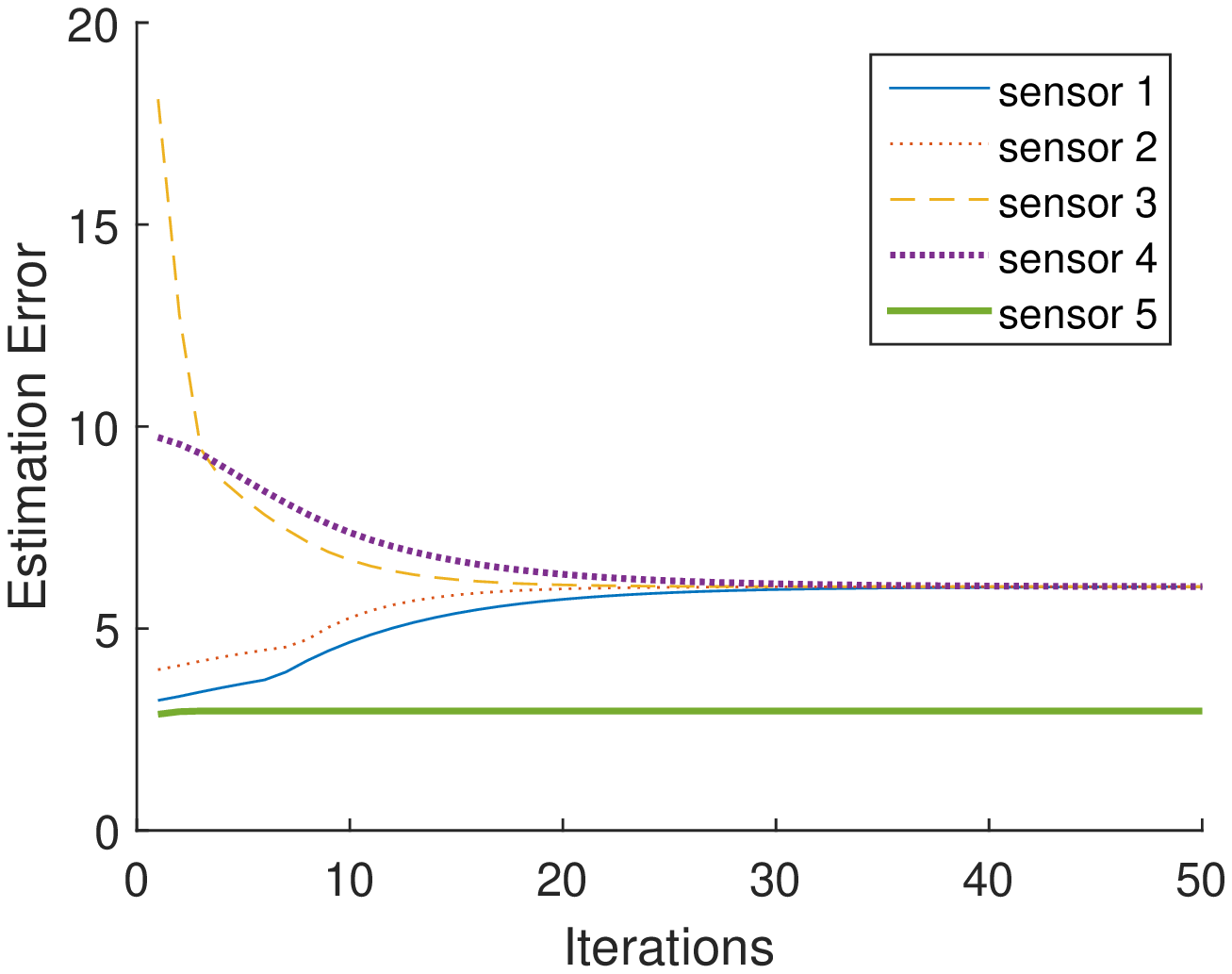}
	\caption{The remote estimation error of each processes.}
	\label{fig:2_cost}
\end{figure}

By running Algorithm~\ref{alg:framework}, we obtain the max-min fair allocation scheme and corresponding average estimation error for each processes. The results of the bandwidth resource allocation for $r_{tot}=2$ are shown in the following Fig.~\ref{fig:2}. It can be seen that the resource allocation converges. Moreover, the average remote estimation errors of all processes, except process~$5$, converge to the same value as shown in Fig.~\ref{fig:2_cost}.

According to Proposition~\ref{proposition: equal performance}, the estimation errors of the unstable processes converge to the same value as expected. Interestingly, although process~$4$ is stable, its estimation error also converges to the same value as those of the unstable processes. Moreover, the communication rate of process~$4$ is the highest among all processes. Meanwhile, process~$5$ as a stable processes, does not occupy any communication bandwidth. Intuitively, process~$4$ is most unpredictable among the five processes in terms of the process noise covariance $Q_i$ while process~$5$ is predictable in the sense that its spectral radius is small and the process noise covariance is very small. Although process 5 is allocated with zero communication budget, its time-averaged estimation error is smaller than other processes with nonzero communication resources.

\begin{figure}[t]
	\centering
	\includegraphics[width=0.35\textwidth]{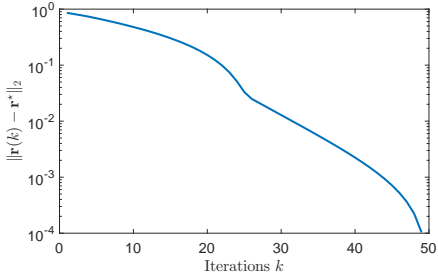}
	\caption{The error of $\bm{r}(k)$ in each iteration.}
	\label{fig:rate}
\end{figure}

As $\underline{r_i}$ is initialized to be small ($0.001$ in the example), the outer loop of the algorithm terminates after one run. Fig.~\ref{fig:rate} shows that the log of the error decays linearly in each iteration in the inner loop as claimed in Theorem~\ref{theorem: convergence to unique point}.

%%%%%%%%%%%%%%%%%%%%%%%%%%%%%%%%%%%%%%%%%%%%%%%%%%%%%%%%%%%%%%%%%%%%%%%%%%%%%%%%%%
\section{CONCLUSION}
In this work, we considered the max-min fairness criterion for resource allocation. We specified the conditions which guarantee the convexity and monotonicity structure of the cost objective with respect to its allocated resources. We used a Lagrangian multiplier method to transform the original problem into a zero-sum game. We developed an algorithm to compute the policy at the equilibrium and proved its convergence. The developed method was applied to a fair sensor scheduling problem. We showed that our formulation includes ``equal" performance of the sensors if it is feasible to have an ``equal-performance" scheduling policy. We utilized the results from the single sensor scheduling problem to efficiently compute the update equations of the subproblems for the decomposed problem.

The coupling between each individual is imposed in terms of the average consumption. In a more strict situation, where the total resource consumption constraint need to be satisfied in each time step, a Markov game setup should be considered. This is out of the framework developed in this work, which is left as future work.

\section*{Appendix}
\subsection{Proof of Theorem~\ref{theorem: equilibrium and optimality}}
We first characterize key properties of an optimal solution of Problem~\ref{problem: resource allocation fairness} and then the equilibrium of the iterative algorithm. Lastly, we combine the two to prove their equivalence. The proof relies on the notation of the normal cone of a set at its boundary. We give its definition here for completeness.
\begin{definition}~\cite{nagurney2012projected}
	Given a set $\mathcal{R}\subset\mathbb{R}^n$ and a point $\bm{r}\in\mathcal{R}$, the normal cone of $\mathcal{R}$ at $\bm{r}$ is 
	\begin{align*}
	N_\mathcal{R}(\bm{r}) := \{ \, \bm{x} :  \bm{x}^\top (\bm{r}'-\bm{r}) \leq 0, \,\forall \bm{r}'\in\mathcal{R} \, \}.
	\end{align*}
\end{definition}

\subsubsection{Characterization of an optimal solution}
The objective function is a pointwise maximum function and is non-differentiable. However, its  subdifferentials exist.
\begin{lemma}\cite{boyd2018subgradients}\label{lemma: subgradient of the pointwise max}
	If the functions $J_i(r_i)$ are subdifferentiable, the subdifferential of $\max_{i\in\mathcal{I}} J_i(r_i)$ is the convex hull of the union the subdifferentials of every $J_j(r_j)$ satisfying $J_j(r_j) = \max_{i\in\mathcal{I}} J_i(r_i)$, i.e.,
	\begin{align*}
	\partial_{\bm{r}} \max_{i\in\mathcal{I}} J_i(r_i) = \textbf{Co}\cup\{ \, \partial_{\bm{r}} J_j(r_j) : \, J_j(r_j) = \max_{i\in\mathcal{I}} J_i(r_i) \, \},
	\end{align*}
	where $\textbf{Co}$ is the convex hull of a set.
\end{lemma}

For subdifferentials, there exists a generalized KKT condition, which involves the normal cone at an optimal solution. The relation between the normal cone and the subdifferential of an optimal solution of Problem~\ref{problem: resource allocation fairness} is as follows.
\begin{lemma}\cite[Theorem 3.33]{ruszczynski2006nonlinear}
	A point $\bm{r}^\star\in\mathcal{R}$ is an optimal solution to Problem~\ref{problem: resource allocation fairness} if and only if
	\begin{align*}
	\bm{0} \in \partial_{\bm{r}} \max_{i\in\mathcal{I}} J_i(r^\star_i) + N_{\mathcal{R}}(\bm{r}^\star),
	\end{align*}
	where $N_{\mathcal{R}}(r)$ is the normal cone of $\mathcal{R}$ at $\bm{r}$ and the plus sign stands for addition in terms of set, i.e., $X+Y=\{\,x+y|: \, x\in X,y\in Y\,\}$ for two sets $X$ and $Y$.
\end{lemma}

Denote the active set of an optimal solution as
\begin{align*}
\mathcal{I}^\star = \{ \, i: \, J_i(r_i^\star) = \max_{j\in\mathcal{I}} J_j(r_j^\star) \, \}.
\end{align*}
There exists one element in the subdifferentials of $\max_{i\in\mathcal{I}} J_i(r^\star_i)$ lying in the space spanned by $\{ \bm{e}_i \}_{i\in\mathcal{I}^\star}$
\begin{lemma}
	There exists $\alpha>0$ such that $\sum_{i\in\mathcal{I}^\star} -\alpha \bm{e}_i\in \partial_{\bm{r}} \max_{i\in\mathcal{I}} J_i(r^\star_i)$, where $\bm{e}_i$ is the unit vector with the $i$-th component being one.
\end{lemma}
\begin{IEEEproof}
	As $\partial_{\bm{r}} \max_{i\in\mathcal{I}} J_i(r^\star_i)$ is the convex hull of $\cup\{ \partial_{\bm{r}} J_i(r^\star_i): \, i\in\mathcal{I}^\star \}$ and the subdifferentials of $J_i(r_i)$ are all negative, the result follows directly.
\end{IEEEproof}

\subsubsection{Characterization of the equilibrium}
For $\bm{r}\in\mathcal{R}$ and $v\in N_\mathcal{R}(\bm{r})$, we have $P_{\mathcal{R}} (\bm{r} + \bm{v}) = \bm{r}$. The equilibrium can thus be characterized with its normal cone.
\begin{lemma}\cite[pp. 12]{nagurney2012projected}
	If $\bm{r}^\star$ is the equilibrium of the discrete-time projected dynamics~\eqref{eq: cost based algorithm}, then
	\begin{align*}
	\bm{J}(\bm{r}^\star) \in N_\mathcal{R}(\bm{r}^\star).
	\end{align*}
\end{lemma}

\subsubsection{Proof of Theorem~\ref{theorem: equilibrium and optimality}}
The direction of $\bm{r}(k)$ in~\eqref{eq: cost based algorithm} forms an acute angle with one of the subdifferentials of the pointwise maximum function as
\begin{align*}
\bm{J}(\bm{r})^\top\Big(-\partial_{\bm{r}} \max_{i\in\mathcal{I}} J_i(r_i) \Big) > 0.
\end{align*}
It is expected that both vectors lie in the normal cone of the equilibrium. The acute angle relation can be further exploited at an optimal solution as follows.
\begin{lemma}\label{lemma:max achieved for active agent}
	If $i\in\mathcal{I}^\star$ and there exists $i'$ such that $J_{i'}(r^\star_{i'})<J_i(r^\star_{i})$, then $r_i^\star = \min\{R,1\}$, which is the maximum allowable resource.
\end{lemma}
\begin{IEEEproof}
	We prove by contradiction. Suppose $i\in\mathcal{I}^\star$ and $r_i^\star < \min\{R,1\}$. Let $r_i = \min\{R,1\}$. In addition, fix one $i'\in\mathcal{I}\backslash\mathcal{I}^\star$, and let $r_{i'} = r_{i'}^\star - (\min\{R,1\} - r^\star_i)$. For other $j\in\mathcal{I}\backslash\{i,i'\}$, let $r_j=r^\star_j$. We can obtain
	\begin{align*}
	&\bm{J}(\bm{r}^\star)^\top (\bm{r} - \bm{r}^\star) \\
	= &J_i(r_i^\star)(\min\{R,1\} - r_i^\star) + J_{i'}(r_{i'}^\star)( r_{i}^\star - \min\{R,1\})\\
	= & ( J_i(r_i^\star) - J_{i'}(r_{i'}^\star) ) (\min\{R,1\} - r_i^\star) > 0.
	\end{align*}
	This contradicts that $\bm{J}(\bm{r}^\star) \in N_\mathcal{R}(\bm{r}^\star)$.
\end{IEEEproof}

Now we are ready to prove Theorem~\ref{theorem: equilibrium and optimality}. We separate the discussion for two cases.
Firstly, if $J_i(r_i^\star) = \max_{j\in\mathcal{I}} J_j(r_j^\star)$ for every $i\in\mathcal{I}$, then
\begin{align}\label{eq: centralized seeking algorithm}
\sum_{i\in\mathcal{I}^\star} J_i(r_i^\star) ( r_i - r_i^\star )
= \sum_{i\in\mathcal{I}} J_i(r_i^\star) ( r_i - r_i^\star ) \leq 0.
\end{align}

Secondly, if there exists $i'$ such that $J_{i'}(r^\star_{i'})<J_i(r^\star_{i})$, then according to Lemma~\ref{lemma:max achieved for active agent}, we have
\begin{align*}
\sum_{i\in\mathcal{I}^\star} J_i(r_i^\star) ( r_i - r_i^\star )
= \sum_{i\in\mathcal{I}^\star} J_i(r_i^\star) ( r_i - \min\{R,1\} ) \leq 0.
\end{align*}
By the definition of $N_{\mathcal{R}}(\bm{r}^\star)$, we have $\sum_{i\in\mathcal{I}^\star} \bm{e}_i\in N_{\mathcal{R}}(\bm{r}^\star)$, where $\bm{e}_i$ is the unit vector with the $i$-th component being one. Since there exists a constant $\alpha>0$ such that 
\begin{align*}
\sum_{i\in\mathcal{I}^\star} -\alpha\bm{e}_i\in&\partial_{\bm{r}} \max_{i\in\mathcal{I}} J_i(r_i) \\
= &\textbf{Co}\cup\{\, \partial_{\bm{r}} J_j(r_j) : \, J_j(r_j) = \max_{i\in\mathcal{I}} J_i(r_i) \, \}.
\end{align*}Therefore,
\begin{align*}
\bm{0} \in \partial_{\bm{r}} \max_{i\in\mathcal{I}} J_i(r_i) +N_{\mathcal{R}}(\bm{r}^\star),
\end{align*}
which completes the proof.

\subsection{Distributed Algorithm}
The algorithm in~\eqref{eq: cost based algorithm} is centralized because of the projection operation. We can develop a distributed version of it by distributing the projection operation. To accomplish such a goal, we take a step backward. Algorithm~\eqref{eq: cost based algorithm} can be perceived as a projected gradient descent algorithm for the following optimization problem
\begin{align}
\min_{\bm{r}} \quad &\sum_{i\in\mathcal{I}} \tilde{J}_i(r_i)\nonumber\\
\text{s.t.} \quad &\sum_{i\in\mathcal{I}} r_i\leq R,~r_i\in[\underline{r_i},\overline{r_i}],\label{eq: equivalent problem}
\end{align}
where
\begin{align*}
\tilde{J}_i(r_i) := \int_{0}^{r_i} -J_i(t)\mathrm{d}t.
\end{align*}
This formulation can be included in a distributed optimization framework as both the objective function and the constraint is written in a separable summation form. We propose one intuitive implementation\footnote{Other implementations such as~\cite{yi2016initialization,nedic2018improved} are also available.}. The dual problem of~\eqref{eq: equivalent problem} can be written as
\begin{align*}
\max_{\lambda\geq0} q(\lambda),
\end{align*}
where $q(\lambda) = \min_{\bm{\underline{r}}\leq\bm{r}\leq\bm{\overline{r}}} \sum_{i\in\mathcal{I}} \Big( \tilde{J}_i(r_i)+\lambda (r_i - R/n) \Big)$. By making $n$ copies of the dual variable $\lambda$, this dual problem can be reformulated as
\begin{align*}
\max_{\bm{\lambda}\geq \bm{0}} \quad &\sum_{i\in\mathcal{I}} q_i(\lambda_i)\\
\text{s.t.} \quad & \lambda_i=\lambda_j \quad \forall i,j\in\mathcal{I},
\end{align*}
where $q_i(\lambda_i)=\min_{\underline{r_i}\leq r_i \leq \overline{r_i}} \tilde{J}_i(r_i)+\lambda_i (r_i - R/n)$. The above constraint can also be relaxed over an undirected graph. Each sensor $i$ can acquire $\lambda_j$ from a subset of $\mathcal{I}$ and the constraint can be compactly written as
\begin{align*}
L\bm{\lambda} = \bm{0},
\end{align*}
where $L$ is a $n\times n$ matrix with its element in the $i$-th row $j$-column being
\begin{align*}
L_{[i,j]} = \begin{cases}
0,& ~\text{if $i$ and $j $ are not connected},\\
-\frac{1}{N_i},&~\text{if $j$ is connceted to $i$},\\
1,&~\text{if}~i=j,
\end{cases}
\end{align*}
where $N_i$ is the number of nodes connecting to node $i$. Consider the following augmented Lagrangian
\begin{align*}
\mathcal{L}(\bm{r},\bm{\lambda}) = \sum_{i\in\mathcal{I}} \Big(\tilde{J}_i(r_i) + \lambda_i(r_i-R/n) \Big) - \bm{\lambda}^\top L \bm{\lambda},
\end{align*}
which is convex in $\bm{r}$ as $J_i(r_i)$ are decreasing and continuous, and concave in $\bm{\lambda}$. The perturbed primal-dual subgradient update can be written as
\begin{align*}
\bm{r}(k+1) &= P_{[\bm{\underline{r}},\bm{\overline{r}}]} \Big(\bm{r}(k) - \epsilon(k)\nabla_{\bm{r}}\mathcal{L}(\bm{r}(k),\hat{\bm{\lambda}}(k+1))\Big)\\
\bm{\lambda}(k+1) &= P_{[\bm{0},+\infty]} \Big(\bm{\lambda}(k) + \epsilon(k)\nabla_{\bm{\lambda}}\mathcal{L}(\hat{\bm{r}}(k+1),\bm{\lambda}(k))\Big),
\end{align*}
where $[\bm{\underline{r}},\bm{\overline{r}}]$ denotes the box constraint for $\bm{r}$ and $[\bm{0},+\infty]$ denotes the positivity constraint for $\bm{\lambda}$.
Note that in the update terms, $\bm{\lambda}(k)$ and $\bm{r}(k)$ has been replaced by two perturbation terms $\hat{\bm{\lambda}}(k+1)$ and $\hat{\bm{r}}(k+1)$. This is adopted to ensure convergence for general convex-concave Lagrangian functions. These perturbation terms $\hat{\bm{\lambda}}(k)$ and $\hat{\bm{r}}(k))$ can be calculated by
\begin{align*}
\hat{\bm{r}}(k+1) &= P_{[\bm{\underline{r}},\bm{\overline{r}}]} \Big(\bm{r}(k) - \hat{\epsilon}(k)\nabla_{\bm{r}}\mathcal{L}(\bm{r}(k),\bm{\lambda}(k))\Big)\\
\hat{\bm{\lambda}}(k+1) &= P_{[\bm{0},+\infty]} \Big(\bm{\lambda}(k) + \hat{\epsilon}(k)\nabla_{\bm{\lambda}}\mathcal{L}(\bm{r}(k),\bm{\lambda}(k))\Big).
\end{align*}
The selection of the step sizes $\epsilon(k)$ and $\hat{\epsilon}(k)$ and the convergence of this perturbed primal-dual subgradient algorithm have been shown in~\cite{kallio1999large}. Since the updating terms
\begin{align*}
\nabla_{\bm{r}}\mathcal{L}(\bm{r}(k),\bm{\lambda}(k)) =& \bm{J}(\bm{r}(k)) + \bm{\lambda}(k)\\
\nabla_{\bm{\lambda}}\mathcal{L}(\bm{r}(k),\bm{\lambda}(k)) =& \bm{r}(k) - R/n - L\bm{\lambda}(k)
\end{align*}
does not require knowledge of $J_i(r_i)$ and $r_i(k)$ of other nodes and the projections can be done for each component, this algorithm is distributed.

\bibliographystyle{IEEETran}
\bibliography{fairbib}
\end{document}